\documentclass[A4paper,
              ]{jacow}
\makeatletter%
	\ifboolexpr{bool{xetex}}
	 {\renewcommand{\Gin@extensions}{.pdf,%
	                    .png,.jpg,.bmp,.pict,.tif,.psd,.mac,.sga,.tga,.gif,%
	                    .eps,.ps,%
	                    }}{}
\makeatother

\ifboolexpr{bool{xetex} or bool{luatex}} % test for XeTeX/LuaTeX
 {}                                      % input encoding is utf8 by default
 {\usepackage[utf8]{inputenc}}           % switch to utf8

\usepackage[USenglish]{babel}			 

\usepackage[final]{pdfpages}
\usepackage{multirow}
\usepackage{ragged2e}

\ifboolexpr{bool{jacowbiblatex}}%
 {%
  \addbibresource{jacow-test.bib}
  \addbibresource{biblatex-examples.bib}
 }{}
\begin{document}

\title{Recent Developments in DEMIRCI for RFQ Design}

\author{
  E. Celebi\textsuperscript{1}\thanks{emre.celebi@cern.ch}, Istanbul Bilgi University, Faculty of Engineering and Natural Sciences, Istanbul, Turkey\\
  G. Turemen\textsuperscript{2}, Ankara University, Department of Physics, Ankara, Turkey\\
  O. Cakir, Ankara University, Department of Physics, Ankara, Turkey\\
 G. Unel, University of California at Irvine, Department of Physics  and Astronomy, Irvine, USA\\
 \textsuperscript{1} also at Bogazici University, Department of Physics, Istanbul, Turkey\\
 \textsuperscript{2} also at TAEK, SANAEM, Ankara, Turkey
}
	
\maketitle

\begin{abstract}
DEMIRCI software aims to aid RFQ design efforts by making the process easy, fast and accurate. In this report, DEMIRCI 8-term potential results are compared with the results provided by other commercially available simulation software. 
 Computed electric fields are compared to the results from simulations of a recently produced 352 MHz RFQ.
 Recent developments like the inclusion of space charge effects in DEMIRCI beam dynamics are also discussed. 
 Moreover, further terms are added to 8-term potential to simulate possible vane production errors. 
 The FEM solver was also improved to mesh the cells with errors. 

\end{abstract}

\section{Introduction}

Radio frequency quadrupole (RFQ) cavities are being used to bunch, focus and accelerate ion-beams at the initial phase of modern linacs. 
The successful operation of an this complex cavity, requires a rapid and realistic calculation of the so called K-T potential \cite{TepilKap}.
The design of the cavity and the simulation of the ion beam within both require  intensive computer simulations with dedicated software.
PARMTEQM and LIDOS are the leading examples to such computer programs that have been developed since 1980s \cite{Parmteq,Lidos}. 
The graphical RFQ design software, DEMIRCI, was born out of the need for a modern and accessible tool \cite{demirint,demirpaper}.
It allows a completely graphical user experience, runs over multiple operating systems and permits, albeit simplistic, beam dynamics calculations  \cite{Dem-recent,Dem-ipac17}. 
This note will focus on the developments made in the last year, such as streamlining the eight term potential coefficients calculation, addition of more terms to calculate the effects of the errors due to construction and assembly, and determination of the RFQ acceptance via beam dynamics.

\section{Eight terms comparison}

The correct determination of the time independent part of the  K-T potential for each 
cell is crucial \cite{TepilKap}. After symmetry, computational time and accuracy considerations,
it has become traditional to calculate the first eight terms (8T) of the full K-T solution  given by:

\begin{eqnarray}
U(r,\theta,z) & = & \frac{V}{2}[\sum_{m=1}^{\infty}r^{2m}A_{0m}\cos(2m\theta)\label{eq:generic-potential}\\
 & + & \sum_{m=0}^{\infty}\sum_{n=1}^{\infty}\cos(nkz)I_{2m}(nkr)A_{nm}\cos(2m\theta)]\nonumber 
\end{eqnarray}

where $r$ and $\theta$ are spherical coordinates for which $z$
represents the beam direction, $V$ is the inter-vane voltage, $k$
is the wave parameter given by $k\equiv2\pi/\lambda\beta$, with $\lambda$
being the RF wavelength and $\beta$ being the speed of the ion relative to the speed of light. 
Additionally, $r_{0}$ is the mean aperture of the vanes, $I_{2m}$ is the modified Bessel function of order $2m$ and
the $A_{nm}$ are the multipole coefficients whose values are to be found depending on the vane geometry. 
DEMIRCI uses the approach of  3 dimensional differential finite element method to obtain 
the potential distribution across the RFQ length and then does a least squares fit to find the multipole coefficients. 
This method was  reported to be as accurate as the image charge method and much faster from computation point of view  \cite{FEM-first}.

To test the validity of the procedure and the correctness of the implementation in DEMIRCI,
the multipole coefficients calculated by DEMIRCI are compared to the ones calculated by other 
groups as found in the literature \cite{FEM-first}. The reference cells for which the coefficients are 
calculated range from low $m$ to high $m$, i.e. from the the entrance of the RFQ to its exit. 
The properties of these reference cells are given in Table \ref{tab:Ref-Cells}. 

\begin{table}
\caption{Definitions of the RFQ Cells Used for Comparison\label{tab:Ref-Cells}}
\centering{}%
\begin{tabular}{|c|c|c|c|}
\hline 
cell\# & $a$ & $m$ & $\ell_{cell}$ \tabularnewline \hline  \hline 
20 & 0.409 & 1.020 & 0.58\tabularnewline \hline 
60 & 0.399  & 1.072 & 0.60\tabularnewline \hline 
100 & 0.392 & 1.111 & 0.68\tabularnewline \hline 
140 & 0.381 & 1.171 & 0.92\tabularnewline \hline 
\end{tabular}
\end{table}

The multipole coefficient calculation results for the reference cells
are shown in Table \ref{tab:Comparison-table}. Each block of 3 rows
contains the calculation results of the multipole coefficients of
the reference cell indicated by the first column. The last column
contains a total error defined as $\epsilon_{tot}=\sqrt{\underset{i}{\sum}(\frac{C_{i}-C_{i}^{p}}{C_{i}^{p}})^{2}}$
where the summation runs over the first 8 coefficients and the reference
values ($C^{P}$) are from PARMTEQM. 

\setlength{\tabcolsep}{5pt}

\begin{table*}
\caption{Comparison of the Multipole Coefficients Calculated by Available Software
\label{tab:Comparison-table}}
\centering{}%
\small{
\begin{tabular}{|c|c||c|c|c|c|c|c|c|c||c|}
\cline{1-1} \cline{3-11} 
cell\# & \multicolumn{1}{c|}{} & $C_{10}$  & $C_{00}$/$a^{2}$  & $C_{11}$  & $C_{01}/a^{6}$  & $C_{30}$  & $C_{20}$  & $C_{31}$  & $C_{21}$  & $\epsilon_{tot} $\tabularnewline
\cline{1-1} \cline{3-11} 
\multicolumn{1}{c}{} & \multicolumn{1}{c|}{} & $A_{10}$ & $A_{01}$ & $A_{12}$ & $A_{03}$ & $A_{30}$ & $A_{21}$ & $A_{32}$ & $A_{23}$ & \tabularnewline
\hline 

 & PARMTEQM & 0.00606 & 5.73007 & 0.05304 & 4.94852 & 0.0 & -0.00003 & 0.0 & 0.00072 & -\tabularnewline \cline{2-11} 
20 & RFQCoef & 0.00601 & 5.74846 & 0.05031 & 4.41814 & 0.0 & -0.00002 & -0.00001 & -0.00077 & 2.325\tabularnewline \cline{2-11} 
 & DEMIRCI & 0.00550 & 5.73009 & 0.06764 & 4.86715 & 0.00000 & 0.00001 & 0.00000 & 0.00002 & 1.561 \tabularnewline \hline  \hline 

 & PARMTEQM & 0.02273 & 5.73981 & 0.20874 & 4.89934 & 0.00000 & 0.00003 & 0.00000 & -0.00031 & -\tabularnewline \cline{2-11} 
60 & RFQCoef & 0.02280 & 5.75573 & 0.21751 & 4.51230 & 0.00000 & -0.00006 & -0.00001 & 0.00402 & 14.322\tabularnewline \cline{2-11} 
 & DEMIRCI & 0.0206315 & 5.74033 & 0.26798 & 4.87713 & 0.00000 & 0.00008 & 0.00000 & 0.00071 & 3.660 \tabularnewline \hline  \hline 

 & PARMTEQM & 0.04307 & 5.74320 & 0.45675 & 4.88818 & 0.0 & 0.00018 & 0.0 & 0.00087 & -\tabularnewline \cline{2-11} 
100 & RFQCoef & 0.04296 & 5.85630 & 0.48357 & 4.03974 & -0.00001 & 0.00022 & -0.00002  & 0.00628 & 6.384\tabularnewline \cline{2-11} 
 & DEMIRCI & 0.03892 & 5.74555 & 0.63027 & 4.88811 & 0.00000 & 0.00029 & 0.00000 & 0.00320 & 2.779 \tabularnewline \hline 
  
 & PARMTEQM & 0.09684 & 5.74616 & 1.40462 & 4.74544 & 0.0000 & 0.00076 & 0.0000 & -0.04081 & -\tabularnewline \cline{2-11} 
140 & RFQCoef & 0.09662  & 5.75831 & 1.35976 & 3.84250 & 0.00000  & 0.00096 & -0.00015  & -0.09994 & 1.791\tabularnewline \cline{2-11} 
 & DEMIRCI & 0.08684 & 5.75535 & 2.57745 & 4.86834 & 0.00001 & 0.00157 & 0.00005 & 0.01996 & 2.016\tabularnewline
\hline 
\end{tabular}
}
\end{table*}

\begin{figure}[!htb]
\centering \includegraphics[clip,width=1\columnwidth]{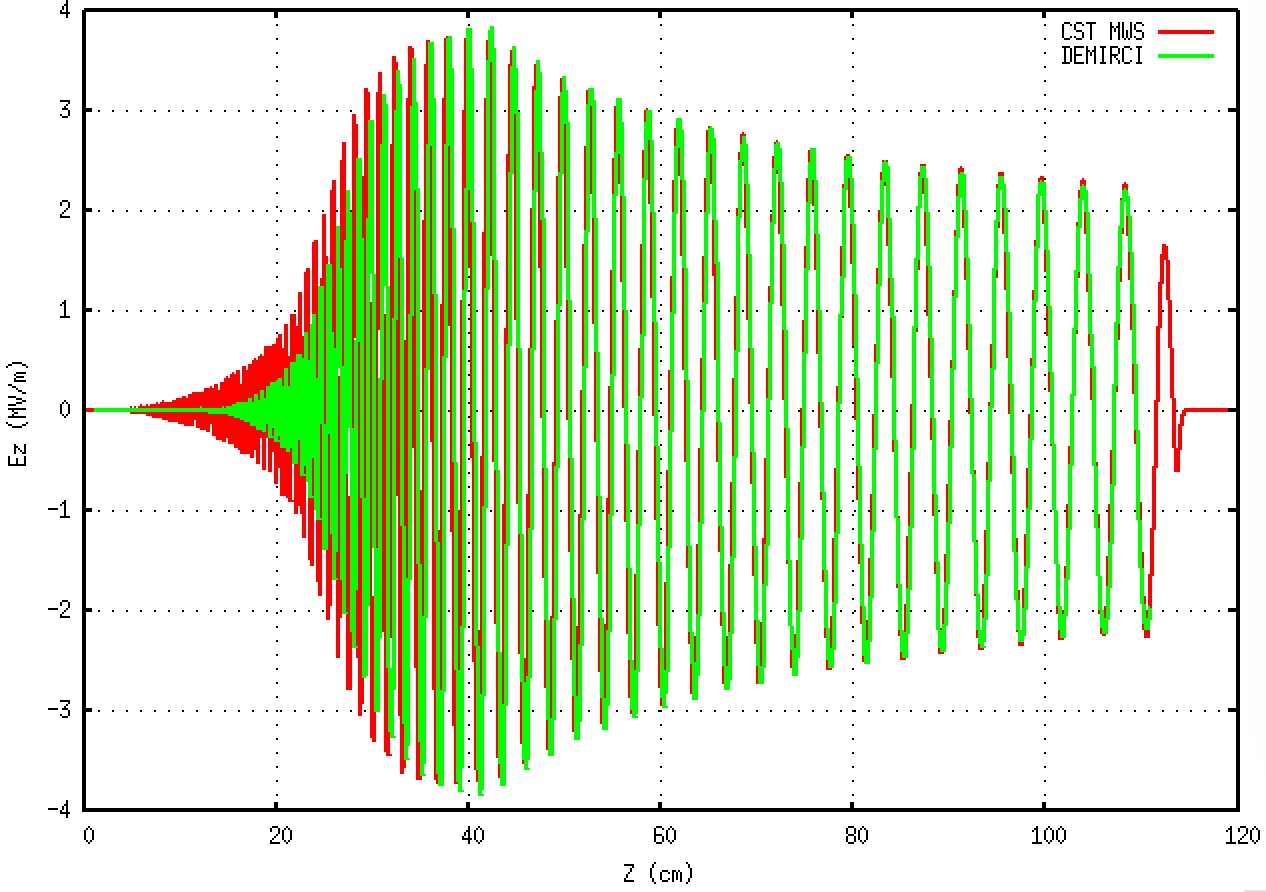} 
\caption{Electric field in the longitudinal direction as calculated by CST and DEMIRCI for the same RFQ. }
\label{dmr-cst} 
\end{figure}

By simply comparing the total error, given in the last column of Table \ref{tab:Comparison-table} and defined relative the PARMTEQM solutions, one can see that coefficients obtained from DEMIRCI calculations are close to those from the other two softwares.
Using the calculated coefficients the 8T potential can be formed and electric field components can be obtained. The Fig. \ref{dmr-cst} contains a  comparison between the $z$ components of the electric field as calculated by CST and DEMIRCI for the same RFQ operating at 352 MHz \cite{demirint}. As one can observe, the exit region (although DEMIRCI does not contain the Crandall cells) and the overall envelope is matching between the two programs, however the input part has a slight mismatch which is being investigated.

Furthermore Demirci-Fem results compared with electrostatic simulation done with CST MWS for a given RFQ cell.
The cell used in the following comparison has a minimum bore radius (a) of 0.309 cm, cell modulation (m) of 1.631 and the cell length of 1.94 cm.

\begin{figure}[!htb]
\centering \includegraphics[clip,width=1\columnwidth]{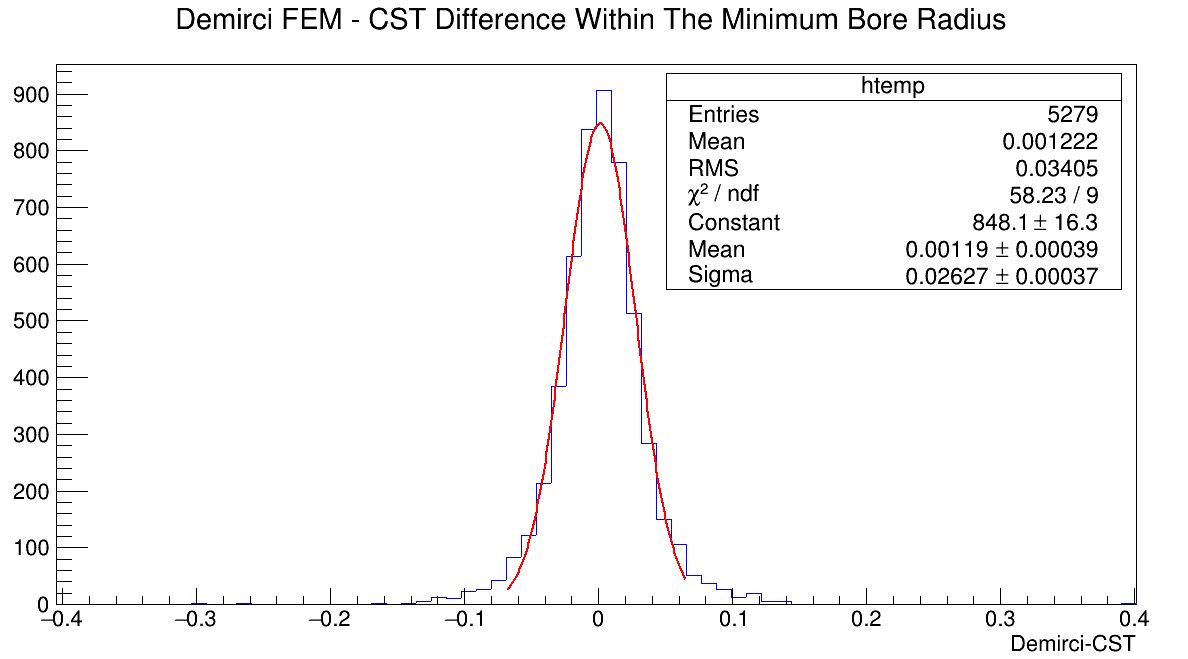} 
\caption{Histogram of potential value differences between CST electrostatic solution and Demirci-Fem solution.}
\label{dmr-cst-cell} 
\end{figure}

Vane profile was produced with Demirci and imported into to CST where it is used to construct the cell geometry.
After electrostatic solution was achieved, values at the mesh node positions produced by Demirci-Fem are scored.
In Fig. \ref{dmr-cst-cell} the difference between CST and Demirci-Fem values are histogrammed. Considering applied voltage between the vanes (200 Volts); it is safe to say that CST and Demirci-Fem results agree with an error less than 0.1\%.

\section {space charge effects in DEMIRCI}
Space charge (SC) effects are handled in a similar way to the 8T potential calculation method. For each cell segment, the vane potentials are given as zero and the poisson equation is solved numerically. However this time, the difference equation becomes an inhomogeneous one due to the charge induced by the macro-particles representing the beam. The potential obtained in this step is to be added to the previously calculated 8T K-T potential and should be used in beam dynamics calculations. At each time step, the SC potential has to be re-calculated for the updated positions of the macro-particles. This is a time (CPU) consuming process and currently, it is being both debugged and speeded up.

\section{Additional terms in K-T potential }
DEMIRCI is also capable of simulating possible vane production and assembly errors. These errors lead to the breakdown of the rotational quadrupole symmetry of the RFQ cavity. Putting back the terms originally cancelled by this symmetry, the equation \ref{eq:generic-potential} becomes:
\begin{eqnarray}
 U_{\epsilon} (r,\theta,z) =   \frac{V}{2} [ \quad \sum_{m=1}^{\infty}r^{m}[A_{0m}\cos m\theta + B_{0m}\sin m\theta ] \label{eq:realistic-potential}\\
   \qquad \qquad  \qquad +  \sum_{m=0}^{\infty}\sum_{n=1}^{\infty}\cos nkz I_{m}(nkr)[A_{nm}\cos m\theta  \nonumber \\ 
   \qquad \qquad \qquad  \qquad \quad  + B_{nm}\sin m\theta ] \quad ] \nonumber 
\end{eqnarray}

where the index $m$ doesn't anymore need to be an even integer and the coefficients $ B_{0m}$ and $ B_{nm}$ are to be determined using the boundary conditions. The FEM solver implementation in DEMIRCI was also improved to handle RFQ cells with errors. Instead of only solving quadrant one, all four quadrants are meshed and the resulting nodes are added to the overall stiffness matrix which is inverted using the conjugate gradient method. 

\section{Acceptance Calculation}
\begin{figure}[!htb]
\centering \includegraphics[clip,width=1\columnwidth]{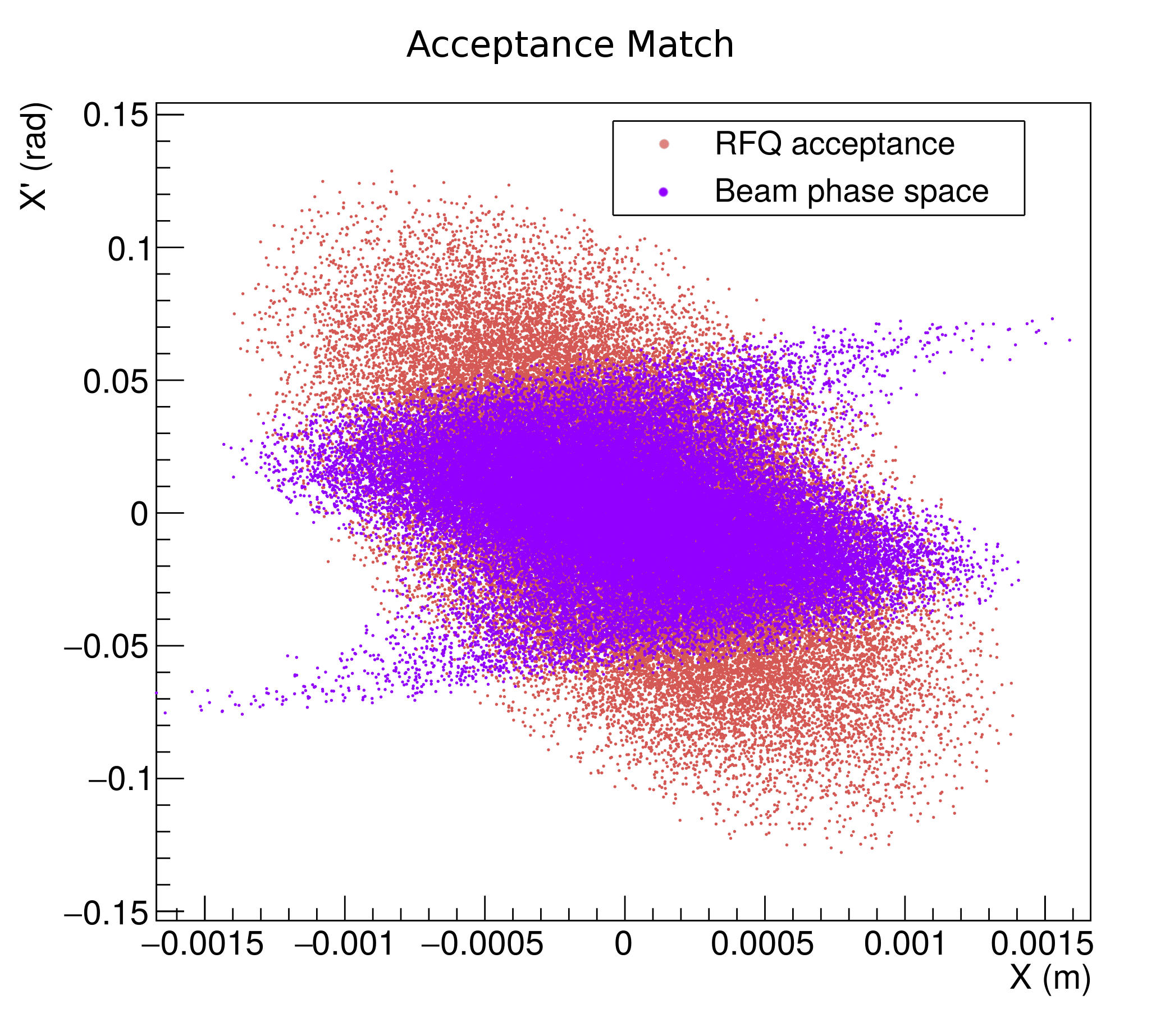} \caption{RFQ beam acceptance and incoming beam's phase space, superimposed.}
\label{acceptance} 
\end{figure}

The acceptance of the RFQ under design, is estimated using a very simple method: a large number of particles are created at the entrance of the RFQ with a uniform distribution in a large portion of the phase space. The particles are then moved through the RFQ using the beam dynamics under the 8T potential as discussed previously. The initial phase space positions of the particles that survive the trip and exit the RFQ, constitute the RFQ acceptance as shown in Fig.\ref{acceptance}. Based on this information the phase space parameters of the RFQ "matched" acceptance ellipse can be found: 
$\alpha_{m}$, $\beta_{m}$ and $\gamma_{m}$ . 
Then the mismatch ($M$) with any incoming beam with known (measured) parameters can be found as:

\begin{equation}
M=\sqrt{1+\frac{{\Delta+\sqrt{(\Delta+4)\Delta}}}{2}}-1
\end{equation}

where $\Delta=\Delta_{\alpha}^{2}-\Delta_{\beta}\Delta_{\gamma}$ and a delta with subscript represents the difference between the beam's Twiss parameters and the "matched" RFQ values, e.g. $ \Delta_{\alpha}=\alpha-\alpha_{m}$.

\section{Outlook}
DEMIRCI is being developed continuously to match the requirements of both national and international RFQ designer communities. 
Its graphical user interface will also contain the interaction mechanisms with other tools to design the ion chamber and the low energy beam
transport magnets. 3D visualization and other aids are in the to do list as well. Comparing DEMIRCI results to those of the PARMTEQM, the state of the art software of the field, is essential to improve its simulation capabilities and to make it more accurate. The procurement of this software is in progress.

\section{Acknowledgment}

This project has been supported by TUBITAK with project numbers 114F106 and 117F143.

\end{document}